\documentclass[superscriptaddress,aps,prx,twocolumn,showpacs,nofootinbib,longbibliography,notitlepage,floatfix]{revtex4-2}
\usepackage{etex}
\usepackage{amsmath,amssymb,amsthm}
\usepackage[ruled,vlined]{algorithm2e}
\usepackage{hyperref}
\hypersetup{colorlinks=true,citecolor=blue,urlcolor=blue}
\usepackage[pdftex]{graphicx}
\usepackage{times,txfonts}
\usepackage{braket}
\usepackage{color}
\usepackage{amsmath,blkarray}
\usepackage{mathtools}
\usepackage{ulem}
\usepackage{latexsym}
\usepackage{tabularx, booktabs}
\usepackage{graphics,epstopdf}
\usepackage{graphicx}
\usepackage{float}
\usepackage{amsfonts}
\usepackage{tikz}
\usepackage{color,soul}

\newcommand{\red}{\color{red}}
\newcommand{\be}{\begin{equation}}
\newcommand{\ee}{\end{equation}}
\newcommand{\ba}{\begin{eqnarray}}
\newcommand{\ea}{\end{eqnarray}}

\usepackage{multirow}
\usepackage{appendix}
\usepackage{url}
\usepackage{bm}

\begin{document}

\title{Digitized-counterdiabatic quantum approximate optimization algorithm}

\author{P. Chandarana}
\thanks{These two authors contributed equally.}
\affiliation{Department of Physical Chemistry, University of the Basque Country UPV/EHU, Apartado 644, 48080 Bilbao, Spain}

\author{N. N. Hegade} 
\thanks{These two authors contributed equally.}
\affiliation{International Center of Quantum Artificial Intelligence for Science and Technology (QuArtist) \\ and Department of Physics, Shanghai University, 200444 Shanghai, China}

\author{K. Paul}
\email{koushikpal09@gmail.com}
\affiliation{International Center of Quantum Artificial Intelligence for Science and Technology (QuArtist) \\ and Department of Physics, Shanghai University, 200444 Shanghai, China}

\author{F. Albarr\'an-Arriagada}
\affiliation{International Center of Quantum Artificial Intelligence for Science and Technology (QuArtist) \\ and Department of Physics, Shanghai University, 200444 Shanghai, China}

\author{E. Solano}
\email{enr.solano@gmail.com}
\affiliation{Department of Physical Chemistry, University of the Basque Country UPV/EHU, Apartado 644, 48080 Bilbao, Spain}
\affiliation{International Center of Quantum Artificial Intelligence for Science and Technology (QuArtist) \\ and Department of Physics, Shanghai University, 200444 Shanghai, China}

\affiliation{IKERBASQUE, Basque Foundation for Science, Plaza Euskadi 5, 48009 Bilbao, Spain}
\affiliation{Kipu Quantum, Kurwenalstrasse 1, 80804 Munich, Germany}

\author{A. del Campo}
\affiliation{Department  of  Physics  and  Materials  Science,  University  of  Luxembourg,  L-1511  Luxembourg, Luxembourg}
\affiliation{Donostia International Physics Center,  E-20018 San Sebasti\'an, Spain}
\affiliation{Department of Physics, University of Massachusetts, Boston, MA 02125, USA}

\author{Xi Chen}
\email{chenxi1979cn@gmail.com}
\affiliation{Department of Physical Chemistry, University of the Basque Country UPV/EHU, Apartado 644, 48080 Bilbao, Spain}

\begin{abstract}
The quantum approximate optimization algorithm (QAOA) has proved to be an effective classical-quantum algorithm serving multiple purposes, from solving combinatorial optimization problems to finding the ground state of many-body quantum systems. Since QAOA is an ansatz-dependent algorithm, there is always a need to design  ansatze for better optimization. To this end, we propose a digitized version of QAOA enhanced via the use of shortcuts to adiabaticity. Specifically, we use a counterdiabatic (CD) driving term to design a better ansatz, along with the Hamiltonian and mixing terms, enhancing the global performance. We apply our digitized-counterdiabatic QAOA to Ising models, classical optimization problems, and the $P$-spin model, demonstrating that it outperforms standard QAOA in all cases we study.
\end{abstract}

\maketitle

\section{Introduction}
Hybrid classical-quantum algorithms have the potential to unleash a broad set of applications in the quantum computing realm. The challenges involved in realizing fault-tolerant quantum computer have promoted the study of such hybrid algorithms, which proved to be relevant to modern noisy intermediate-scale quantum (NISQ) devices~\cite{qaoa2,Bharti21} with few hundred qubits and limited coherence time. One notable example is that of variational quantum algorithms (VQA), which is implemented by designing variational quantum circuits to minimize the expectation value for a given problem Hamiltonian. 
VQA is advantageous given the fact that preparing a tunable circuit ansatz is found to be difficult on a classical computer. It has already been widely applied in quantum chemistry~\cite{vqeapp1,vqeapp2,vqeapp3,vqeapp4,vqeapp5,vqeapp6}, condensed matter physics~\cite{vqe7,vqe9,vqe10}, solving linear system of equations \cite{vqls}, combinatorial optimization problems \cite{vqeco,vqeco1} and several others \cite{vqf1,vqf2}. Remarkably, one of the early implementations of the VQA was performed using photonic quantum processors~\cite{Peruzzo2014}, which prompted further theoretical progress~\cite{theoryvqe1,theoryvqe2,theoryvqe3,theoryvqe4,theoryvqe5,theoryvqe6}. VQA has been demonstrated in superconducting qubits~\cite{theoryvqe1,vqeapp1,vqeapp4} and trapped ions~\cite{vqeapp6,theoryvqe7,ising6}.

 One compelling outcome of VQA is the development of the quantum approximate optimization algorithm (QAOA)~\cite{qaoa1}, which provides an alternative for solving combinatorial optimization problems using shallow quantum circuits with classically optimized parameters. In the past few years, there has been a rapid development in QAOA-based techniques that have been applied not only for solving conventional optimization problems like MaxCut but also for solving ground state problems in different physical systems \cite{Willsch_2020,ps9,ising6}. Improved versions of QAOA, like ADAPT-QAOA~\cite{adapt} and Digital-Analog QAOA~\cite{daqaoa} have also been reported recently. Like any combinatorial optimization problem, QAOA depends on optimizing a cost function to obtain the desired optimal state corresponding to a $p$-level parametrized quantum circuit. In addition, the choice of the approximate trial state, from which the cost function is obtained, is crucial to the success of the QAOA algorithm. Generally, this is done by using quantum adiabatic algorithms (QAA) which produce near-optimal results for large $p$ which is not suitable for current NISQ devices. Moreover, due to the requirement of large $p$, the cost of classical optimization increases and the algorithms suffer from the problem of vanishing gradients and local minima~\cite{McClean2018,Cerezo_2021,vqe8}. 

Several studies have been reported in past few years showing that high fidelity quantum states can be prepared by assisting QAA with additional driving interaction~\cite{nar}. These studies establish that for certain problems, the inclusion of additional driving terms can reduce the computational complexity, and with it the circuit depth. These driving terms are usually calculated using methods developed under the umbrella of so-called shortcuts to adiabaticity ~\cite{sta1,sta16}, which have been introduced to improve the traditional quantum adiabatic processes, removing the requirement for slow driving~\cite{sta2}. Instances of these methods include counterdiabatic (CD) driving~\cite{sta3,sta4,sta5}, fast-forward approach~\cite{ffsta6,ffsta7}, and invariant-based inverse engineering~\cite{insta8,insta9}. Among them,  CD driving is interesting and has been used to study fast dynamics~\cite{sta9,sta10,sta11,sta12,sta13}, preparation of entangled states~\cite{ensta1,ensta2,ensta3,ghz}, adiabatic quantum computing~\cite{stan1,nar} and quantum annealing~\cite{annsta1,annsta2,ps8}.

In the context of QAOA, the advantage of the introduction of CD driving is twofold. The CD driving decreases the circuit depth, while reducing the number of optimization parameters. On the other hand, it provides a better approximate trial state which is beneficial for finding the optimal target state. In this work, we propose a novel algorithm, digitized-counterdiabatic quantum approximate optimization algorithm (DC-QAOA), which improves the performance of the conventional QAOA using CD driving. In this context, it is worthwhile to mention the work of Ref.~\cite{qa1}, also inspired by CD driving techniques.

% In the following sections, to compare their respective performances, we show the application of DC-QAOA and QAOA to different problems like Ising spin models, Classical optimization problems that include Maxcut and SK model, and the $P$-spin model. The purpose behind studying these problems is to show the competitive performance of this algorithm. We measure the performance of our algorithm by comparing the approximation ratios.
This article is organized as follows. In Sec.~\ref{staqaoa}, we introduce DC-QAOA algorithm and explain it in detail, comparing it with the quantum adiabatic evolution and QAOA. %In Sec.~\ref{isings}, Sec.~\ref{classicop}, Sec.~\ref{pspinsec}, we depict the improvement by using DC-QAOA as compared to QAOA for various models by comparing the approximate ratio ($\mathcal{R}$). 
In the following sections we present a comparative study of the proposed DC-QAOA and the conventional QAOA method in the context of various physical systems. In Sec.~\ref{isings}, we prepared the ground state of three different types of 1D Ising spin models, namely, the Longitudinal field Ising model (LFIM), the transverse field Ising model (TFIM),  and the GHZ state. In Sec.~\ref{classicop}, we studied classical optimization problems such as the MaxCut problem and the Sherrington-Kirkpatrick model, while in  Sec.~\ref{pspinsec}, different variants of $P$-spin model are considered. In doing so, we establish, by comparing the approximation ratios, that the DC-QAOA is advantageous compared to the QAOA for shallow quantum circuits. We conclude with a discussion in Sec.~\ref{concl}.
\section{Digitized counter-diabatic quantum approximate optimization algorithm (DC-QAOA)}\label{staqaoa}
\begin{figure}%[h]
\includegraphics[width=1\linewidth]{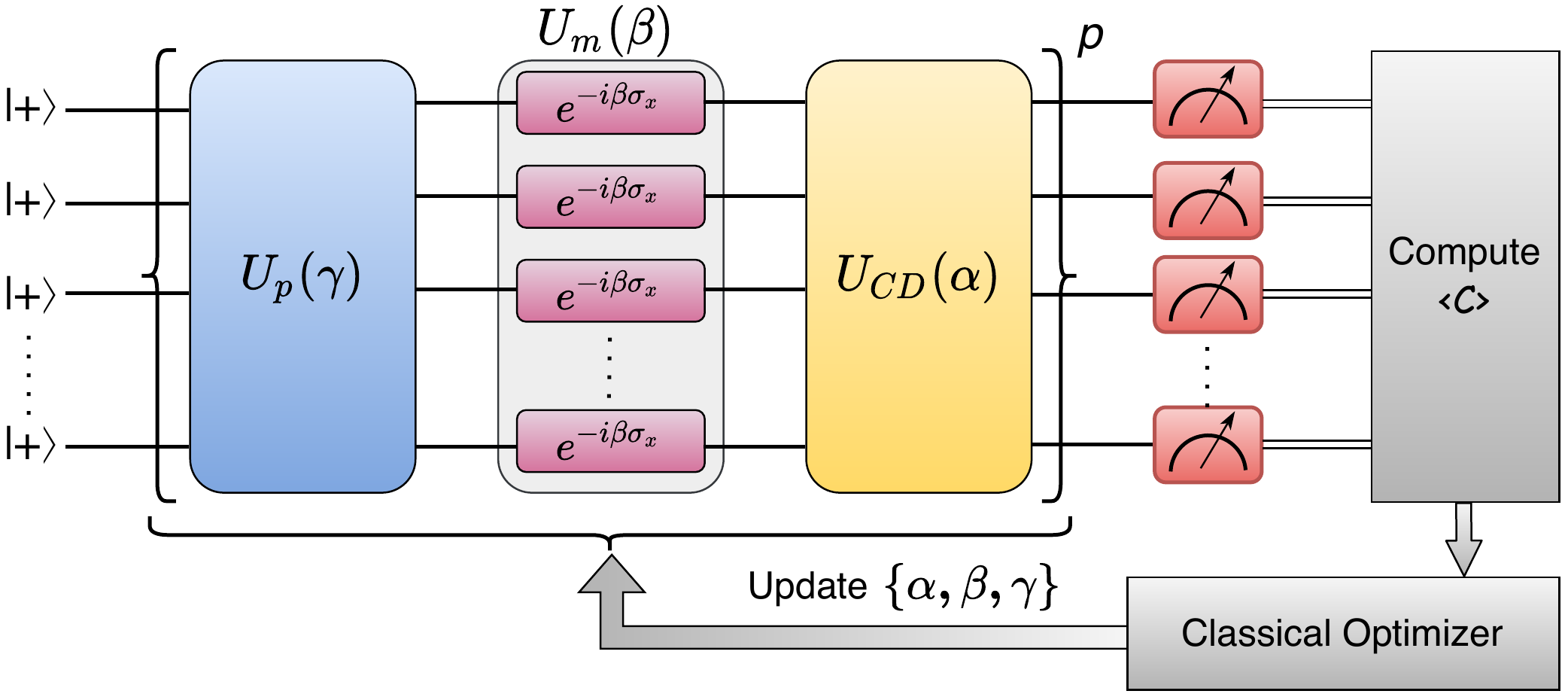} 
    \caption{Schematic diagram with circuit used for DC-QAOA having additional CD term along with the Hamiltonian and mixing terms.}
    \label{circuitcdqaoa}
\end{figure}
%CD driving provides a universal scheme to design shortcuts to adiabaticity and, in theory, it can speed up almost any adiabatic evolution and improve the fidelity of the target state. 
%This fact is particularly important in this study as in QAOA, 
%As such, it is important to prepare a variational trial state which is as close as possible to the actual target state. In general, 
The  conventional QAOA method can be viewed as a combination of two distinct parts: the quantum part consists of a parameterized circuit ansatz, which is in turn complemented by a classical optimization algorithm to determine the parameters that minimize (maximize) a predefined cost function. The circuit ansatz for the quantum part is governed by an annealing Hamiltonian,
%%. 
\begin{equation}
    H_a(t) = ( 1 - \lambda(t))H_{mixer} + \lambda(t)H_{prob},
    \label{anneal}
\end{equation}
%%
%where $\lambda(t)$ is the annealing schedule for $t \in \{ 0, T\}$ and $H_{prob}$ is the system Hamiltonian. This corresponds to the trial state that we try to optimize. The adiabatic circuit ansatz can be designed using the trotterized time evolution operator \cite{trot,Suzuki1976
%{\red where $\lambda(t)\in \{ 0, 1\}$ is the annealing schedule for $t \in \{ 0, T\}$. \st{and} $H_{mixer} = \sum_i h_i \sigma^x_i$ is the mixing Hamiltonian that produces an equal (weighted) superposition state to begin with. The system Hamiltonian  $H_{prob}$ corresponds to the trial state that we try to optimize. In continuous annealing, the system evolves from the eigenstate of $H_{mixer}$ to the eigenstate of $H_{prob}$ through adiabatic evolution. The corresponding digital adiabatic circuit ansatz can be designed using the trotterized time evolution operator \cite{trot,Suzuki1976}}
 where $\lambda(t)\in [ 0, 1]$ is the annealing schedule for $t \in [0, T]$. $H_{mixer} = \sum_i h_i \sigma^x_i$ is the mixing Hamiltonian that produces an equal (weighted) superposition state  in the computational basis to begin with, whereas the desired final state is the ground state (or an eigenstate) of $H_{prob}$. In continuous annealing, the system evolves from the eigenstate of $H_{mixer}$ to the eigenstate of $H_{prob}$ through adiabatic evolution. The corresponding digital adiabatic circuit ansatz can be designed using the trotterized time evolution operator \cite{trot,Suzuki1976}
\begin{equation}
U(0, T) \approx \prod_{j=1}^{p} \prod_{m=1}^M \exp \left\{-i H_{m}(j \Delta t) \Delta t\right\},
\label{trott}
\end{equation}
where we consider that $H_a(t)$ can be decomposed into $M$ $k$-local terms, i.e., into terms $H_m(t)$ which have $k$-body interactions at most.
Note that the $U(0,T)$ is a product of $p$ sub-unitaries, each corresponding to an infinitesimal propagation step $\Delta t$. An adiabatic evolution using $U(0,T)$ can always produce an exact target state at the cost of resorting to a large value of $p$. This can be translated to the language of QAOA if one parameterizes $U(0,T)$ as
\begin{equation}
    U(\bm{\gamma}, \bm{\beta}) = U_m{(\beta_p)}U_p{(\gamma_p)}U_m{(\beta_{p-1})}U_p{(\gamma_{p-1})} \dots U_m{(\beta_1)}U_p{(\gamma_1)},
\end{equation}
where the evolution operators are $ U_m{(\beta_p)} = \exp{(-i \beta_p H_{mixer})}$ and $ U_p{(\gamma_p}) = \exp{(-i \gamma_p H_{prob})}$. Here,  the annealing schedule is characterized by the discrete set of parameters  $\{ \beta_p,\beta_{p-1},\dots, \beta_1 \}$ and  $\{ \gamma_p,\gamma_{p-1},\dots, \gamma_1 \}$. $( \bm{\gamma}, \bm{\beta})$ defines a $2p$ parameter space that corresponds to the depth of the circuit ansatz and the cost function $F(\bm{\gamma},\bm{\beta})$ is optimized classically to obtain an optimal parameter set $(\bm{\gamma}^*,\bm{\beta}^*)$, which produces the desired target state,  i.e., $\ket{\psi(\bm{\gamma}^*,\bm{\beta}^*})$. Note that in most cases this target state is chosen to be the ground state of $H_{prob}$.

\begin{figure*}[t]
\includegraphics[width=1\linewidth]{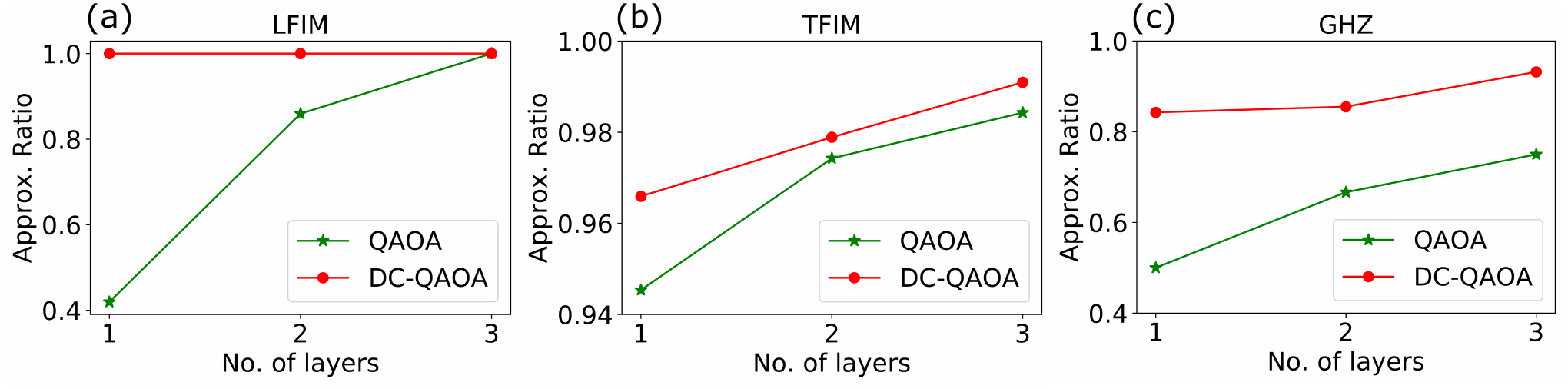}
\caption{Comparison of approximation ratios ($\mathcal{R}$) as a function of number of layers ($p$) for three representative cases of Ising spin model. Green lines show results of QAOA whereas red lines show results of DC-QAOA. ({\red a}) shows $\mathcal{R}$ variation of LFIM where $J_{ij} =1$, $h_i =1$, and $k_i=0$. ({\red b}) depicts TFIM where $J_{ij} =1$, $h_i =0$, and $k_i=1$, and ({\red c}) shows preparation of GHZ state where $J_{ij} =1$, $h_i =0$, and $k_i=0$. System size for all the cases was kept to $L=12$ qubits.}
    \label{isingfig}
\end{figure*}

As the case of adiabatic evolution, QAOA requires large $p$ to obtain a near-optimal trial state, even with the assistance of the classical optimizer. In addition, the realization of $U(\bm{\beta}, \bm{\gamma})$ for an interacting many-body system for large $p$ becomes inefficient due to  the large number of gates involved. %Instead, we opt for the CD approach to obtain a better trial state. 
In DC-QAOA, we focus on improving the quantum part of QAOA, by adding a variational parameter  in each step, i.e.,
\begin{equation}
    U(\bm{\gamma},\bm{\beta}) \rightarrow U(\bm{\gamma},\bm{\beta},\bm{\alpha}), \;\;\; F(\bm{\gamma},\bm{\beta}) \rightarrow F(\bm{\gamma},\bm{\beta}, \bm{\alpha}),
\end{equation}
%%
%The resulting circuit ansatz is shown in Fig.~\ref{circuitcdqaoa}. 
The application of another parameter decreases the size of $p$ drastically. This additional parameter can be quantified as the inclusion of the CD driving term in the problem Hamiltonian. The resulting circuit ansatz is shown in Fig.~\ref{circuitcdqaoa}.

In general, CD driving amounts to using an additional control Hamiltonian in Eq.~\eqref{anneal}, required for suppressing non-adiabatic transitions \cite{del_Campo_2013,sta3,sta4,sta5}.  This is especially effective for many-body systems with tightly spaced eigenstates. CD driving comes at a cost,  as it generally involves nonlocal many-body interactions, and their exact specification of the CD Hamiltonian term requires access to the spectral properties of the driven system \cite{sta3,sta5,sta9}.  As a way out, variational approximations have been proposed to obtain the CD terms \cite{ensta1,Saberi14,qa2}.
In this context, one can use the adiabatic gauge potential for finding an approximate CD driving without spectral information of the system \cite{qa2,agp}. %As a further advance, DC-QAOA only requires the operator form of the CD driving combined with the additional set of parameters $\bm{\alpha}$. DC-QAOA is also more flexible in regards to the boundary conditions compared to the CD evolution which permits the application of the driving term even for one step only. Moreover, the operators can be chosen heuristically and according to the requirement of the system which is being studied. In the following sections, a pool of operators is defined using a second-order expansion of the nested commutator (NC) ansatz \cite{Claeys19}, from which we chose based on the success probability of our algorithm.  We transform each term in the expansion into separate operators and put them into the operator pool  $A =\{e^{-i\alpha A_t}\}_{t=1}^T$, where $A_1$, $A_2$,...  shows each of the constituent terms up to $A_T$, with $T$ being the maximum number of components in the expansion. The addition of a new free parameter through an operator will increase the degrees of freedom, making it possible to reach broader parts of the Hilbert space of the Hamiltonian with a lower circuit depth than in QAOA

 In the following sections, a pool of CD operators is defined using the nested commutator approach of the adiabatic gauge potential, provided by Claeys et. al. \cite{Claeys19},
\begin{equation}
    A_{\lambda}^{(l)} = i \sum_{k = 1}^l \alpha_k(t) \underbrace{[H_{a},[H_{a},......[H_{a},}_{2k-1}\partial_{\lambda} H_{a}]]] .
    \label{gauge}
\end{equation}
Here we considered up to the second order in the expansion of the nested commutator, $l=2$, which gives rise to an operator pool $A = \{\sigma^y, \sigma^z \sigma^y, \sigma^y \sigma^z,  \sigma^x \sigma^y,  \sigma^y  \sigma^x\}$, including solely local and two-body interactions. Note that the choice of this operator pool depends on $H_a(t)$ and may contain other operators depending on the problem Hamiltonian. However, $A$ contains every possible CD operator that can be derived from the problem Hamiltonians, used in the present study. We chose the CD term as a combination of these operators for each system based on the success probability of the algorithm.  For instance, the local CD driving term provides better success probability in the case of the transverse-field Ising model and $P$-spin model, however not suitable for solving the MaxCut Hamiltonian.  %We transform each term in the expansion into separate evolution operators and put them into the operator pool  $A =\{e^{-i\alpha A_t}\}_{t=1}^T$, where $A_1$, $A_2$,...  shows each of the constituent terms up to $A_T$, with $T$ being the maximum number of components in the expansion.
The CD coefficients $\alpha_k$ are transformed into the additional variational parameter associated with the CD driving. The addition of such a new free parameter increases the degrees of freedom, making it possible to reach broader parts of the Hilbert space of the Hamiltonian with a lower circuit depth than in QAOA. Furthermore, as DC-QAOA only requires the operator form of the CD driving combined with the additional set of parameters $\bm{\alpha}$, it eliminates the requirement of complex calculation of the CD coefficients. DC-QAOA is also more flexible in regards to the boundary conditions compared to the CD evolution which permits the application of the driving term even for one step only. Moreover, the operators can be chosen heuristically and according to the requirement of the system which is being studied.

Although there are several ways to define the cost function, we opt for the most convenient one which is the energy expectation value of the problem Hamiltonian calculated for the trial wave function, 
\begin{equation}
     F(\bm{\gamma},\bm{\beta},\bm{\alpha}) = \bra{\psi(\bm{\gamma},\bm{\beta},\bm{\alpha})}H_{prob}\ket{\psi(\bm{\gamma},\bm{\beta},\bm{\alpha})},
     \label{fsta}
\end{equation}
where $\psi(\bm{\gamma},\bm{\beta},\bm{\alpha})$ represents the approximate trial state produced by the digitized CD ansatz. The efficiency of our algorithm can be measured in terms of the approximation ratio, given by,
\begin{equation}
    \mathcal{R} = \frac{F(\bm{\gamma},\bm{\beta},\bm{\alpha})}{{E_0}},
    %R = F/e
    \label{aresta}
\end{equation}
 where $E_0$ is the ground state energy of the system. 

Classical optimization techniques are an integral part of variational algorithms, which helps to find the optimal parameters that minimize the cost function. Our work mainly considers two optimization techniques, namely Momentum Optimizer and Adagrad Optimizer, which are specific examples of stochastic gradient descent (SGD) algorithms. Momentum Optimizer is a variant of SGD in which a momentum term is added along with the gradient descent. The prime purpose of the momentum term is to increase the parameter update rate when gradients are in the same direction and decrease the update rate when gradients point in a different direction~\cite{momentumoriginal}. On the other hand, Adagrad Optimizer’s main purpose is to change the update rate based on the past descent results~\cite{adagradoriginal}. Adagrad has shown great improvements in the robustness of SGD~\cite{adagrad11}. These two classical optimization techniques work pretty well for the cases we consider. This is because these optimization routines have proven faster convergence than gradient descent. Moreover, some of the cases we study involve a large Hilbert space, which may lead to local minima in the energy landscape. In the presence of steep gradients, the use of these techniques proves beneficial. This problem dependence of the performance is shared with other optimization routines such as Nesterov Momentum, Adam, and AdaMax.  An overview and comparison about challenges faced by the different types of gradient descent optimization, can be found in Ref.~\cite{grad}.
 
\section{Ising spin Models}\label{isings}
 1D quantum Ising spin chains are the manifestation of the simplest many-body systems that are widely studied in existing quantum processors. Numerous computational problems can be mapped to finding the ground state of the Ising-like Hamiltonians, which makes it suitable for benchmarking various quantum algorithms. The general form of the Hamiltonian of 1D Ising spin model is given by,
\begin{equation}
     H_{prob}(\sigma) = -\displaystyle\sum_{<i,j>}J_{ij}\sigma_i^z\sigma_j^z - \displaystyle\sum_{i}h_i\sigma_i^z - \displaystyle\sum_{i}k_i\sigma_i^x,   \label{isingh}
\end{equation}
where $\sigma_i^\delta$ denotes the Pauli matrices at the $i$th site, and $<i,j>$ corresponds to the nearest-neighbor interaction with strength $J_{ij}$. The on-site interaction terms $h_i$ and $k_i$ represent the longitudinal and transverse fields, respectively. We consider the periodic boundary conditions so that our model describes a ring of interacting spins~\cite{ising1,ising2,ising3}. Note that three special cases can be retrieved from Eq.~\eqref{isingh}: i) longitudinal field Ising model (LFIM) when $k_i=0$, ii) transverse field Ising model (TFIM) when $h_i=0$, and iii) a special case when both $k_i=0$ and $h_i=0$, for which the resulting ground state of $H_{prob}$ is the highly entangled Greenberger-Horne-Zeilinger (GHZ) state~\cite{ghz1,ghz2,ghz3,ghz4}. For simplicity, we choose the system to be homogeneous i.e., $J_{ij}=J$ as well as $h_{i} =h_z$ and $k_{i} = h_x$. To prepare an equal superposition of the qubits, as a input of the circuit ansatz, the mixing Hamiltonian is chosen as $H_{mixer} = \sum_{i}\sigma_i^x$. 
To implement DC-QAOA, as mentioned in Sec.~\ref{staqaoa}, along with the problem and mixer Hamiltonian, we include the CD term to define the circuit ansatz. %This is done by defining a pool of CD operators, $A_t = \{\sigma^y, \sigma^z \sigma^y, \sigma^y \sigma^z,  \sigma^x \sigma^y,  \sigma^y  \sigma^x\}$ from which we heuristically choose one. 
 The CD operator is chosen heuristically from the operator pool $A$. For instance, in the case of LFIM, the ground state is ferromagnetic and constitutes a large energy gap with the first excited state for the chosen interaction strengths. In such cases, the local driving term $A_t=\sum_{i}\sigma^y_i$ can produce the ground state. On the other hand, the ground state of TFIM is closely spaced with the nearby excited states, which makes the local driving term insufficient. Similarly, the local driving term is also not suitable for GHZ state \cite{del_Campo_2013}.  Instead, the second-order term $A_t=\sum_{i}\sigma^z_i\sigma^y_{i+1}$ is more likely to produce a better result. The unitary operator that represents the CD part of the circuit ansatz is given by,
\begin{equation}
  U_{CD}(\alpha) = \prod_{j=1}^{L}e^{-i\alpha A_t^q}, 
  \label{ualpha}
\end{equation}
where $A_t^q$ represents the respective $q$-local CD operator chosen from the CD pool $A$. For instance, if $q=1$ then $A_t^q = \{A_t\}_j$ and if $q=2$, then $A_t^q=\{A_t\}_{j,j+1}$. The circuit is designed using the gate model of quantum computing whereas the classical optimization is the stochastic gradient descent method. Fig.~\ref{isingfig} depicts the improvement obtained by DC-QAOA over traditional QAOA. In the simulation, we study a $12$-qubit system, for which we compute $\mathcal{R}$ for different $p$ values. For LFIM, as shown in Fig.~\ref{isingfig}a, $\mathcal{R}=1$ even for $p=1$ with DC-QAOA which constitutes considerable improvement over QAOA, that requires $p=3$ to achieve unit $\mathcal{R}$. Hence, for Fig.~\ref{isingfig}a, the number of variational parameters required to achieve unit $\mathcal{R}=1$ for DC-QAOA is $3p = 3$ whereas for QAOA it is $2p=6$. We also see that, for a lower number of layers i.e., $p=1,2,3$, DC-QAOA converges faster to the unit $\mathcal{R}$ compared to QAOA.  Furthermore, while DC-QAOA shows better convergence at lower depths, for the TFIM and the GHZ states, the exact ground state can only be achieved with $p \geq L/2$ layers. %(see Fig.~\ref{isingfig}b and \ref{isingfig}c){\red . 
This effect can be attributed to the Lieb-Robinson bound \cite{nachtergaele2010liebrobinson,scipost} which forces the circuits for TFIM and GHZ state to scale linearly with the system size in order to achieve unit $\mathcal{R}$.

To compare the resource requirements, both classical and quantum, one can inspect two crucial elements of these methods. In the case of systems with nearest-neighbor interactions, the increase in circuit depth per layer by adding a CD term will be constant, and it depends on the CD term chosen. The circuit depth can be quantified as $d \times p$, and the CD driving increases it to $(d +d_{cd} ) \times p$, where $d_{cd}$ represents the increment in depth per layer. For LFIM, the CD term $\sigma_y$ gives $d_{cd}=1$, whereas for TFIM $d_{cd}=4$ for $\sigma_z\sigma_y$. On the other hand, the increase in parameter space due to the CD term is always from $2p$ to $3p$, making DC-QAOA advantageous specifically for low $p$ values. In the limit of large $p$, the performance of  QAOA and DC-QAOA becomes comparable for fixed system size.
% Another way to compare the performances is to make a comparison of the number of layers implemented to achieve a specific approximation ratio ($\mathcal{R}$). For example, in Fig.~\ref{isingfig}b, $\mathcal{R}\approx0.97$ can be achieved using an ansatz with $p=1$ using DC-QAOA whereas QAOA needs $p=2$ to reach $\mathcal{R}=0.97$ which makes DC-QAOA advantageous.    }

\section{Classical Optimization problems}\label{classicop}
Thus far, we have discussed the applications of DC-QAOA for finding the ground state of the Ising model and preparing entangled states. Combinatorial optimization problems are another set of problems that can be encoded in the ground state of a quantum Hamiltonian, diagonal in the computational basis. Here we discuss the application of DC-QAOA for solving combinatorial optimization problems, where the main objective is to find the optimal solution for a given classical cost function. MaxCut is one fundamental combinatorial optimization problem that has been solved using QAOA.

\begin{figure}
    \centering
    \includegraphics[width =\linewidth]{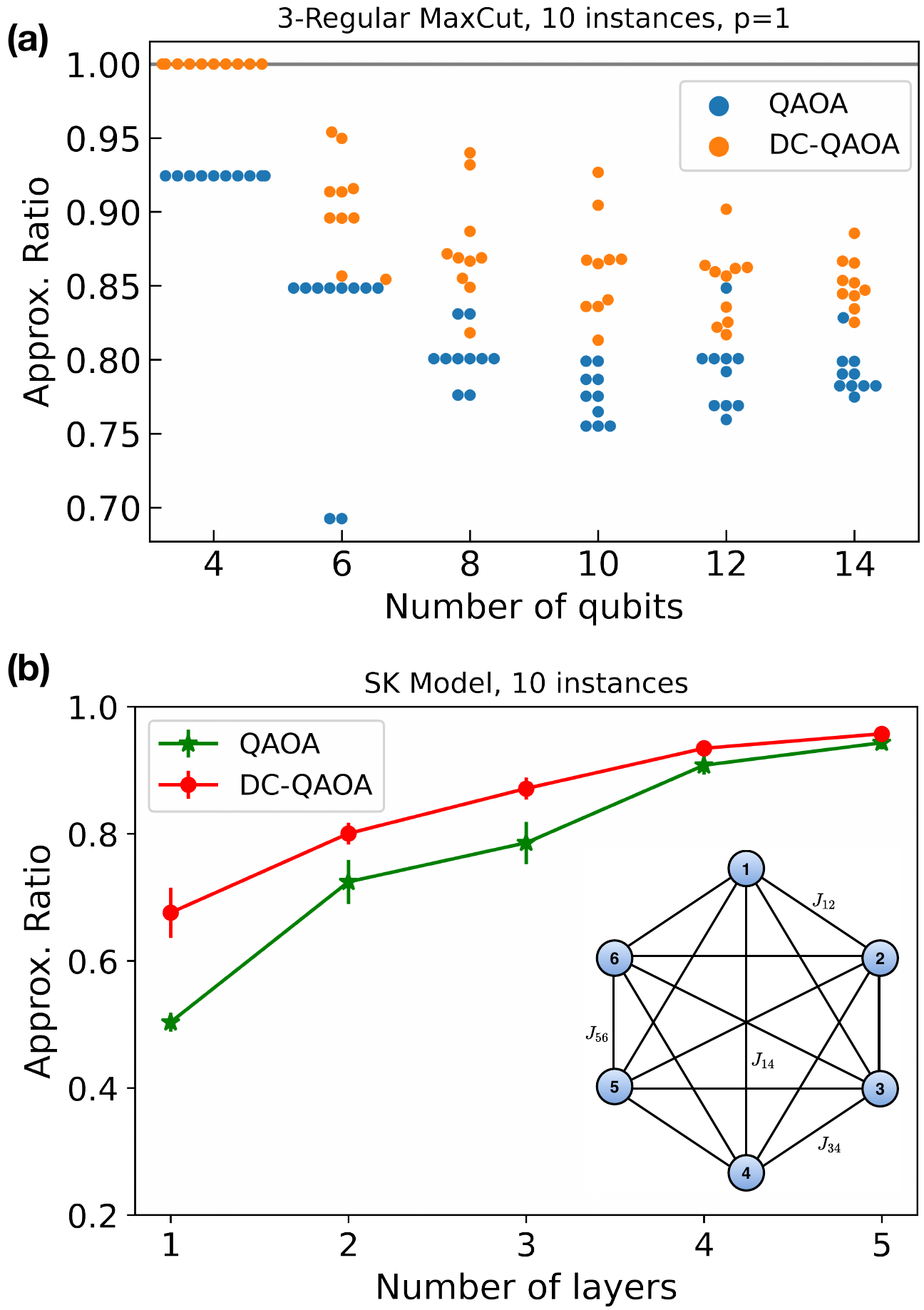} 
    \caption{Comparison of  approximation ratios obtained for different graph size using DC-QAOA and QAOA.  
    \textbf{(a)} Unweighted 3-regular MaxCut for a randomly chosen 10 instances.
    \textbf{(b)} Approximation ratio vs number of layers ($p$) for SK model with 6-qubits (vertices) is depicted. Green line and red line show the values of QAOA and DC-QAOA respectively. On the right-bottom, a graph of 6 qubits with all-to-all connectivity is also shown. The results were obtained by considering 10 different randomly chosen instances of $J_{ij}$ values. Error bars represents the standard error.} 
    \label{MaxCut_SK}
\end{figure}

\begin{figure*}
\includegraphics[width=0.8\linewidth]{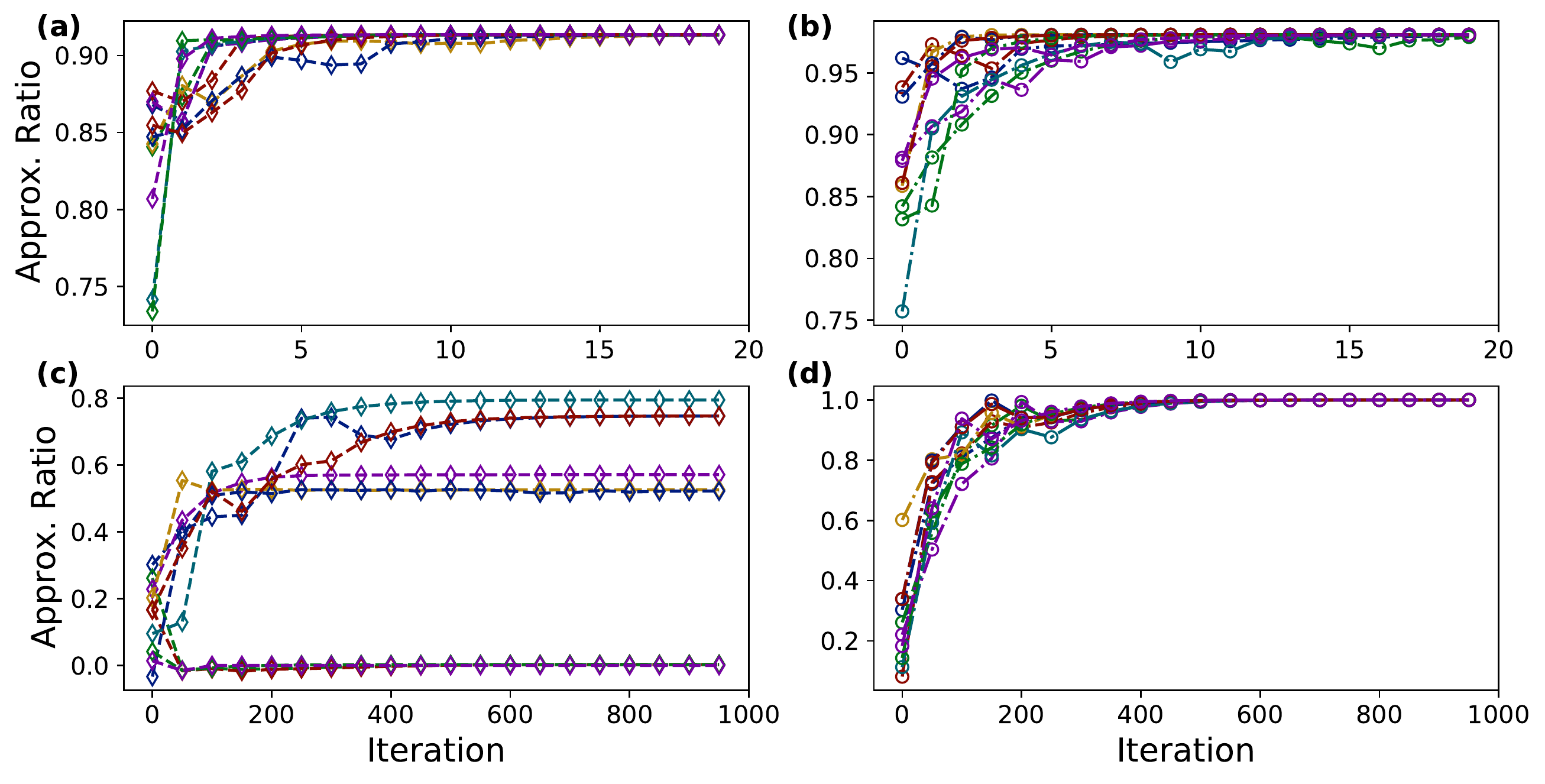}
\caption{Comparison of approximation ratio ($\mathcal{R}$) with respect to number of iterations for $P=4$ and $p=1$. Panels ({\red a}) and ({\red b}) show the QAOA and DC-QAOA results for $h=1$, respectively. The case with $h=0$ is shown in panels ({\red c}) and ({\red d}) for QAOA and DC-QAOA results, respectively. The system size is $L=6$.  Each of the 10 random initial parameters chosen is represented by a color line.}
    \label{pspinfig}
\end{figure*}

For the MaxCut problem, let us consider a graph $G = (V,E)$, where $V$ and $E$ being the vertex set and edge set respectively. We consider a classical cost function $C(z)$ defined on binary strings $\boldsymbol{z}=\left(z_{1}, z_{2}, \ldots, z_{n}\right)$, and aim at  separating the vertices into two sets so that the number of edges cut by $C(z)$ is {\it maximized}. This maximizes the classical cost function
\begin{align}
 C(z) =\frac{1}{2}\sum_{(i,j) \in E} w_{ij} (1-z_i z_j),
\label{cost_c}
\end{align}
where $w_{ij}$ represents the edge weight between vertices $i$ and $j$. Depending on the sets that the vertices of each edge are in after the cut, binary values (either $0$ or $1$) are assigned to variables $z_i$ and $z_j$ corresponding to respective vertices. This situation can be encoded in the ground state of the problem Hamiltonian by mapping the binary variables to Pauli operators
\begin{equation}
    H(\sigma) = \displaystyle\sum_{(i,j) \in E}J_{ij}\sigma_i^z\sigma_j^z.
    \label{skh}
\end{equation}
Note that Eq.~\eqref{skh} also belongs to the Ising class and is equivalent to Eq.~\eqref{isingh} for GHZ states if only nearest neighbor interaction is considered, which is the case of the $2$-regular MaxCut. Here, to verify the performance of our algorithm, we consider unweighted ($w_{ij} = J_{ij} =1$) $3$-regular MaxCut problem, with each vertex connected to three other vertices. 
The CD operator pool can be obtained from the NC expansion and is given by 
$A = \{\sigma^z\sigma^y , \sigma^y\sigma^z\}$. In Fig.~\ref{MaxCut_SK} (a), the approximation ratio $\mathcal{R}$ %$R = {C(z)_{obtained}}/{C(z)_{exact}}$ 
for different graph sizes with up to 14 vertices (qubits) are shown for a single layer ($p=1$). We notice that for small graph sizes, say $4$ qubits), DC-QAOA is superior as it reaches unit $\mathcal{R}$. However, for a bigger graph $\mathcal{R}$ decreases gradually while exceeding the performance of QAOA. Although this can be improved for $p>1$ but for large depth DC-QAOA, the number of parameters for each step scales as $3p$, so the landscape of the cost function most likely has a complicated form, and we expect to see the problem of vanishing gradients (Barren plateau). A detailed analysis is needed for $p>1$ DC-QAOA, which we leave for future work.

Interestingly, if $J_{ij}$ is chosen as random all-to-all two-body interactions, Eq.~\eqref{skh} represents the so-called Sherrington-Kirkpatrick (SK)~model. SK model is a classical spin model proposed by Sherrington and Kirkpatrick \cite{sk1,sk75} where $J_{ij}$ are interaction terms such that $J = \{ \forall J_{ij}\}$ has zero mean and unit variance. For instance, they can be randomly chosen from the set $J=\{-1,1\}$ with probability $1/2$. The SK model is interesting for DC-QAOA as it can be studied as a combinatorial search problem on a complete graph.  QAOA on the SK model has been extensively studied recently~\cite{skqaoa1,skqaoa2}. Here, ten different instances of $J_{ij}$ values are considered in a system of  $L=6$ spins.  Note that the couplings  $J_{ij}$ are non-uniform and the CD term depends on the choice of $J_{ij}$. As this model involves similar interactions to that in the MaxCut problem, we chose the CD term from the same operator pool, $A = \{ \sigma^z \sigma^y , \sigma^y  \sigma^z\}$, calculated from the nested commutator ansatz. In fact, the CD-term chosen for the SK model is $A_t=J_{ij}  \sigma_i^z \sigma_j^y$, where the operators are applied to all the sites due to its all-to-all connectivity. %as the SK model is an all-to-all connected model.} %These operators were chosen from the operator pool A={σ^y σ^z,σ^z σ^y }, calculated from the nested commutator ansatz. }%As this model has similar interaction as MaxCut, we chose the CD operator as $A_t = \{ \sigma^z \sigma^y , \sigma^y  \sigma^z\}$ in Eq.~\eqref{ualpha}.

In Fig.~\ref{MaxCut_SK}b, the approximation ratio ($\mathcal{R}$) is shown with respect to a varying number of layers ($p$). We observe that $\mathcal{R}$ is higher for DC-QAOA as compared to QAOA and that as the number of layers increases DC-QAOA and QAOA start to converge to the same value. This shows that DC-QAOA is efficient for instances where the circuit ansatz is low-layered. In fact, for low layers, although not giving the exact ground state DC-QAOA gives significantly enhanced $\mathcal{R}$. This could be advantageous as we can find optimal parameters which could be used as initial parameters for high-layered QAOA. 

\section{$P$-spin Model}\label{pspinsec}
As a final benchmark, we consider the $P$-spin model, which is a long-range exactly-solvable fully-connected model   ~\cite{ps1,ps2,ps3,ps4}. The system Hamiltonian reads
\begin{equation}
     H = -\frac{1}{L^{P-1}}\left(\sum_{i=1}^{L}\sigma^z_i\right)^{P} - h\sum_{i=1}^{L}\sigma^x_i
    \label{pspin}
\end{equation}
While the ground-state of Hamiltonian   (\ref{pspin}) is trivial, the presence of a quantum phase transition makes its preparation challenging by quantum annealing~\cite{ps1}. For $P=2$ this Hamiltonian exhibits a second-order phase transition whereas a first-order phase transition occurs for $P \geq 3$, closing the energy gap exponentially with increasing system size. This has motivated proposals to change the first-order phase into second-order phase transition by making the Hamiltonian non-stoquastic~\cite{ps5,ps6}. The nature of the ground state also depends on $P$. For odd $P$ the ground state is non-degenerate, while for even $P$ it has a two-fold degeneracy with $Z_2$ symmetry, which makes the choice of the CD operator difficult~\cite{ps8}.
\begin{figure}
    \centering
    \includegraphics[width=1\linewidth]{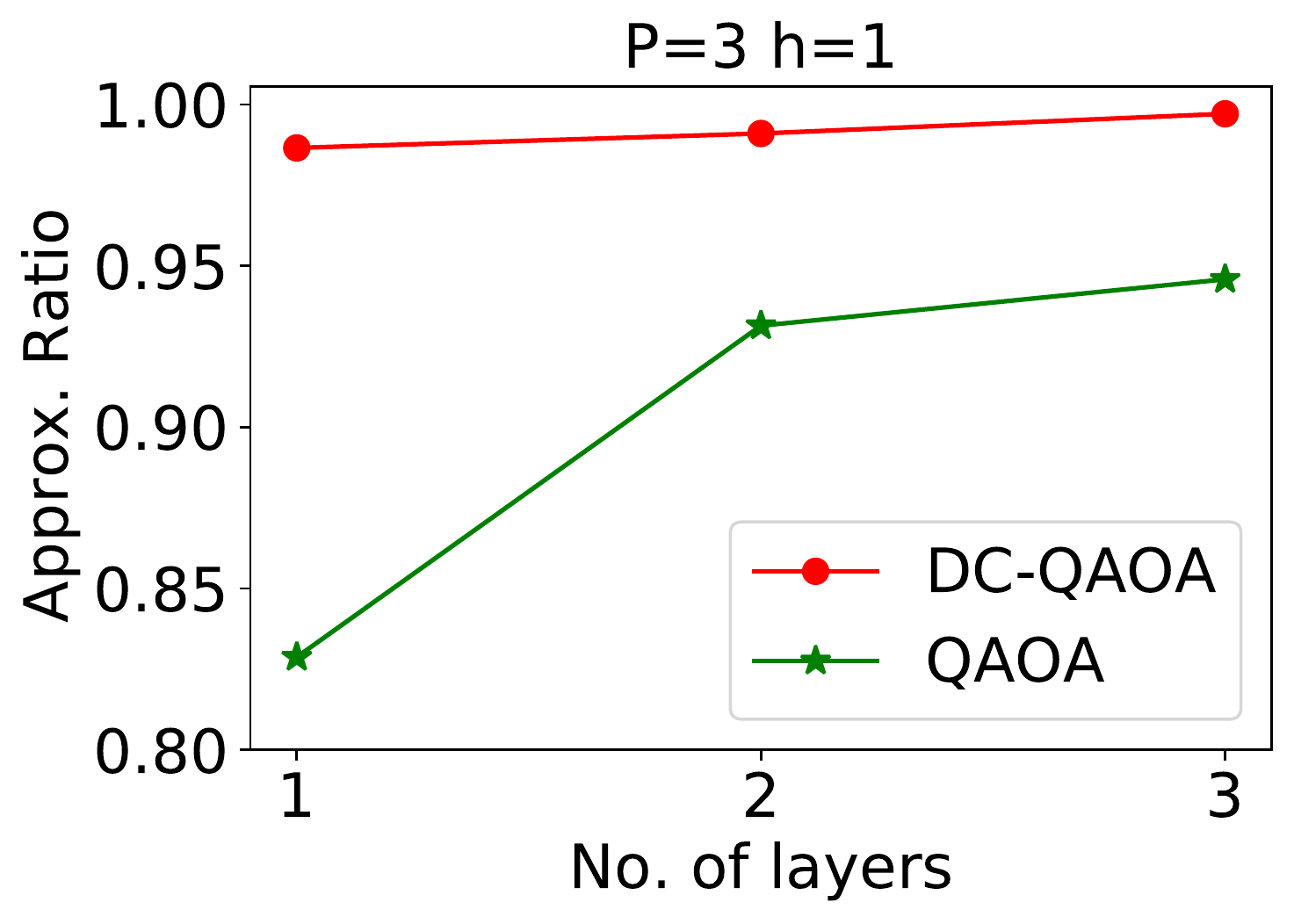}
    \caption{Approximation ratio ($\mathcal{R}$) as a function of number of layers ($p=1,2,3$) for $P=3$ and $h=1$. Green and red lines show  the average results obtained from 10 random parameter initialization for QAOA and DC-QAOA, respectively. Standard deviations are of the order of $10^{-2}$.  }
    \label{p3h1}
\end{figure}
We study DC-QAOA in a $6$ qubit $P$-spin model for the nontrivial case of $h\neq0$ using local CD operator $A_t=\sum_{i}\sigma^y_i$. QAOA and DC-QAOA are compared for three different cases: $P=3$ , $ h = 1$ and $P=4$, $h=\{0,1\}$ respectively. Fig.~\ref{pspinfig} and Fig.~\ref{p3h1} shows the advantage obtained by DC-QAOA for $P=4$ and $P=3$ respectively. For $P=4$,  $\mathcal{R}$ as a function of number of iterations is shown for $p=1$ for $10$ random parameter initialization. We observe that for a finite number of iterations, DC-QAOA shows higher $\mathcal{R}$ values as compared to QAOA for both $h=0$ and $h=1$. It is evident that, in the case of $h=0$, QAOA is highly dependent on the choice of initial parameters and lands into local minima in some instances. By contrast, DC-QAOA shows unit $\mathcal{R}$ for every instance. For $P=3, h=1$, $\mathcal{R}$ is shown as a function of number of layers ($p=1,2,3$) for $10$ random initial parameters. As expected, for DC-QAOA, $\mathcal{R}$ values end up close to unity even for $p=1$ and $\mathcal{R}$ values increase as the number of layers increases. However, this is not surprising for $P = 3$ as the ground state is a product state making it favorable for the local CD operator. The more intriguing case is in Fig.~\ref{pspinfig}b, where the approximation ratio reaches close to unity for $p=1$ even when the ground state is degenerate. This occurs simply because the trial state converges to a particular one of the two due to the local CD driving. This is in contrast with QAOA, which does not achieve the target state for $p=1$ in any case.

\section{Discussion and Conclusion}\label{concl}
We have introduced a quantum algorithm leveraging the strengths of shortcuts to adiabaticity for quantum approximate optimization algorithms. Specifically, we have formulated a variant of QAOA using CD driving, called DC-QAOA, and established its enhanced performance over QAOA in finding ground states of different models. We benchmark our algorithm by considering various examples, starting with Ising spin models, preparing entangled states, classical optimization problems like MaxCut and SK model and, the P-spin model. Including the CD term to the circuit ansatz, the performance of the QAOA algorithm is enhanced. Results reveal that for low-layered circuits, DC-QAOA converges to the ground state faster than state-of-the-art QAOA. Thus, adding a new free parameter in the form of a gate chosen from a predefined set (CD term) increases the performance of the algorithm for shorter circuit depths. Thus, DC-QAOA turns out to be a preferable algorithm for circuits of shorter depth.

In conclusion, DC-QAOA outperforms QAOA for all the models we have studied. For high-depth circuits,  DC-QAOA can be applied for initial layers only to enhance the performance of the standard QAOA. An interesting prospect would be to use the resulting optimal parameters from low depth DC-QAOA as the initial parameters of a high-depth QAOA in order to obtain the minima of the cost function efficiently. %Also, an interesting prospect would be to check if the resulting optimal parameters from low depth DC-QAOA can be used as the initial parameters of standard QAOA with high depth circuit}. 
Our work shows that implementing principles of shortcuts-to-adiabaticity  to enhance quantum algorithms has both fundamental and practical importance. The experimental realization of DC-QAOA on real hardware offers an exciting prospect for further progress. 

{\it Note:} As we finished this work, we learned about the recent preprint devoted to QAOA assisted by CD ~\cite{wurtz2021counterdiabaticity}.

\section*{Acknowledgments}
This work is supported by NSFC (12075145), STCSM (2019SHZDZX01-ZX04), Program for Eastern Scholar, Basque Government IT986-16, Spanish Government PGC2018-095113-B-I00 (MCIU/AEI/FEDER, UE), projects QMiCS (820505) and OpenSuperQ (820363) of EU Flagship on Quantum Technologies, EU FET Open Grants Quromorphic (828826) and EPIQUS (899368). X. C. acknowledges the Ram\'on y Cajal program (RYC-2017-22482).

\bibliography{main.bib}

\end{document}